\documentclass[lettersize,journal]{IEEEtran}
\usepackage{amsmath,amsfonts}
\usepackage{algorithmic}
\usepackage{algorithm}
\usepackage{array}
\usepackage[font=normalsize,labelfont=sf,textfont=sf]{subcaption}
\usepackage[caption=false,font=normalsize,labelfont=sf,textfont=sf]{subfig}
\usepackage{textcomp}
\usepackage{stfloats}
\usepackage{url}
\usepackage{verbatim}
\usepackage{graphicx}
\usepackage{cite}
\usepackage{bm}
\usepackage{booktabs}
\usepackage{graphicx}
\begin{document}
\title{Density Adaptive Attention-based Speech Network: Enhancing Feature Understanding for Mental Health Disorders}

\author{
Georgios Ioannides\textsuperscript{1,2,4,*,$\dagger$}\thanks{$\dagger$These authors contributed equally to this work.}\thanks{*Work does not relate to position at Amazon.}
\and
Adrian Kieback$^{1,\dagger}$
\and
Aman Chadha$^{1,3,4,*}$
\and
Aaron Elkins$^{1}$\\
\IEEEauthorblockA{
    $^1$James Silberrad Brown Center for Artificial Intelligence\quad $^2$Carnegie Mellon University\\   $^3$Stanford University\quad
    $^4$Amazon GenAI
}\\
\IEEEauthorblockA{
    \texttt{gioannid@alumni.cmu.edu, adrian.kieback@gmail.com, hi@aman.ai, aelkins@sdsu.edu}
}
}



\maketitle

\begin{abstract}
Speech-based depression detection poses significant challenges for automated detection due to its unique manifestation across individuals and data scarcity. Addressing these challenges, we introduce DAAMAudioCNNLSTM and DAAMAudioTransformer, two parameter efficient and explainable models for audio feature extraction and depression detection. DAAMAudioCNNLSTM features a novel CNN-LSTM framework with multi-head Density Adaptive Attention Mechanism (DAAM), focusing dynamically on informative speech segments. DAAMAudioTransformer, leveraging a transformer encoder in place of the CNN-LSTM architecture, incorporates the same DAAM module for enhanced attention and interpretability. These approaches not only enhance detection robustness and interpretability but also achieve state-of-the-art performance: DAAMAudioCNNLSTM with an F1 macro score of 0.702 and DAAMAudioTransformer with an F1 macro score of 0.72 on the DAIC-WOZ dataset, without reliance on supplementary information such as vowel positions and speaker information during training/validation as in previous approaches. Both models' significant explainability and efficiency in leveraging speech signals for depression detection represent a leap towards more reliable, clinically useful diagnostic tools, promising advancements in speech and mental health care. To foster further research in this domain, we make our code publicly available\footnote{\url{https://github.com/sdsuai/depression_audio_processing}}
\end{abstract}

\begin{IEEEkeywords}
Speech Depression Detection, Explainable and Interpretable Artificial Intelligence
\end{IEEEkeywords}

\section{Introduction}
Acoustic modeling has become pivotal in identifying psychological and emotional conditions \cite{Cummins, FLINT1993309}. Research has consistently highlighted the efficacy of speech processing in the automated, unbiased identification of psychiatric disorders, notably Major Depressive Disorder (MDD) \cite{pillai}. A diverse array of diagnostic indicators and computational strategies have been introduced for depression detection, each presenting its unique strengths and challenges \cite{Rejaibi, Shen2022, Chlasta2019}. Included among these are spectral \cite{Zhao2015, Liu2017}, prosodic \cite{Williamson2019}, vocal timbre \cite{Dubagunta2019}, and speech production characteristics \cite{Huang2020}, as well as advanced computational techniques like data augmentation \cite{Srivastava2014}, ensemble methods \cite{Ringeval2019}, transfer learning \cite{chen2022speechformerhierarchicalefficientframework}, and self-supervised pre-training \cite{chen2022speechformerhierarchicalefficientframework}. Furthermore, the analysis of features related to the identity of the speaker has been explored for the detection of depression \cite{Wang2023, Bailey2021}, though their claimed results are not reproducible.

The realm of mental health diagnostics, especially for conditions like depression and MDD, faces significant challenges due to the scarcity and imbalance of publicly available datasets \cite{Li2019SpeechDF}. Traditional machine learning and deep learning approaches often struggle with these limitations, leading to models that may not generalize well across the diverse spectrum of depressive symptoms and severities encountered in clinical practice. This issue is compounded by the fact that depression manifests uniquely across individuals \cite{Farina2018}, requiring diagnostic systems to be highly sensitive to subtle and varied vocal biomarkers. Previous approaches have often been constrained by their computational complexity and the extensive number of parameters, rendering them impractical for real-time applications \cite{chen2022speechformerhierarchicalefficientframework}. These models, while powerful, demand significant computational resources, limiting their deployment in scenarios where quick analysis is crucial. Additionally, many existing methods are not end-to-end and/or rely on supplementary information, such as vowel classification \cite{Feng2022} and speaker-specific details \cite{Wang2023}, which are not always readily available or necessitate labor-intensive manual labeling.

According to the World Health Organization (WHO), depression is a pervasive mental health disorder that significantly impacts individuals and societies worldwide \cite{WHO2023a}. It is characterized by a persistent low mood, loss of interest or pleasure in activities, and a range of physical and emotional symptoms that can severely affect a person's ability to function in daily life \cite{AmericanPsychiatricAssociation2013}. Depression is not just a temporary change in mood or a sign of weakness; it is a serious medical condition that requires understanding and treatment \cite{NIMH2024}. In 2019, approximately 280 million people worldwide were living with depression, highlighting its status as a major public health concern. The prevalence of depression varies by country, with the United States reporting a rate of 8.3\% among its adult population in 2021 \cite{MajorDepression2023, WHO2017, WHO2023a}.

The integration of Convolutional Neural Networks (CNNs) and Long Short-Term Memory (LSTM) networks has been a promising direction in addressing these challenges, leveraging CNNs' ability to extract rich spectral features and LSTMs' proficiency in capturing temporal dynamics \cite{Abdel-Hamid2014}. However, the effectiveness of these models can be significantly hindered by data scarcity and imbalance. To mitigate these issues, this work introduces DAAMAudioCNNLSTM, a novel lightweight hybrid CNN-LSTM model ($\approx 280K$ parameters), enhanced with a multi-head Density Adaptive Attention Mechanism (DAAM) \cite{Ioannides2024}. DAAM is especially well-suited for handling the challenges of depression datasets. By using Gaussian distributions, it can dynamically target the most informative parts of speech data, enabling the model to prioritize essential features even with limited and imbalanced data.

In addition to DAAMAudioCNNLSTM, we introduce the DAAMAudioTransformer, a novel transformer-based model that utilizes a similar attention mechanism but replaces the CNN-LSTM architecture with a transformer encoder. Transformer models, characterized by their self-attention mechanisms, have demonstrated remarkable success in various natural language processing tasks and are well-suited for capturing long-range dependencies and contextual relationships within the data \cite{Vaswani2017}. The DAAMAudioTransformer, with 1.1M parameters, achieves the state-of-the-art F1 macro score of 0.72 on the DAIC-WOZ dataset, offering a powerful alternative to traditional architectures.

Moreover, DAAM offers significant explainability benefits, a critical consideration in the healthcare industry. By providing clear insights into which features and segments of the speech signal were deemed most important for a given diagnosis, healthcare professionals can gain a deeper understanding of the model's decision-making process. This transparency is essential for building trust in automated diagnostic systems and facilitating their integration into clinical workflows \cite{Dwivedi2023}. It also opens up new avenues for research, allowing clinicians and researchers to explore the specific speech characteristics most indicative of depression, potentially uncovering novel biomarkers and contributing to a more nuanced understanding of the condition. DAAMAudioCNNLSTM and DAAMAudioTransformer, owing to the Density Adaptive Attention Mechanism, advance the field towards more explainable, trustworthy, and clinically useful diagnostic tools. By focusing on these key areas, this work aims to contribute to the broader goal of improving mental health care through technology, ensuring that patients receive timely, accurate, and personalized diagnoses. In brief, our contributions are as follows:
\begin{itemize}
    \item We introduce the DAAMAudioCNNLSTM, an enhanced and explainable version of the original DepAudioNet, incorporating a novel multi-head DAAM module. This mechanism is designed to transform input audio encodings into highly informative representations. By leveraging learnable parameters, DAAM focuses on the most salient features for the depression detection task, thereby enhancing the model's interpretability and effectiveness.
    
    \item We introduce the DAAMAudioTransformer, a transformer-based model incorporating the DAAM module, achieving a state-of-the-art F1 macro score of 0.72 on the DAIC-WOZ dataset. This model leverages the self-attention mechanisms of transformers to capture long-range dependencies and contextual relationships within the data, providing a robust alternative to traditional architectures \cite{Vaswani2017}.

    \item We establish new state-of-the-art benchmarks for the DAIC-WOZ \cite{Gratch2014} dataset by training DAAMAudioCNNLSTM and DAAMAudioTransformer to identify and aggregate the most informative features within the speech signals for accurately detecting depression. Our approaches surpass all previous end-to-end state-of-the-art models that rely solely on label information, without the need for additional data such as speaker information or vowel positions. This achievement underscores the efficiency and robustness of DAAMAudioCNNLSTM and DAAMAudioTransformer in utilizing only audio signals for depression detection, making them valuable for clinicians and researchers alike.

\end{itemize}

\section{Related Work}
Depression detection has been approached from various angles, utilizing different modalities such as text, speech, and visual data. Among the early works, facial expression analysis has shown promise in identifying depressive states. For instance, the system presented by Nazira et al. (2021) \cite{Nazira2021} uses CNNs in conjunction with OpenCV and Haar Cascade Classifiers to analyze facial expressions for depression detection. This method demonstrated an accuracy of 81\% and a recognition rate of 88\%, highlighting the potential of visual data in mental health diagnostics. However, the need for specialized datasets and the difficulty of capturing consistent visual data limit the applicability of these methods in real-world settings.

In contrast, speech-based models offer significant advantages, including non-invasiveness and the ability to capture data in natural conversational settings. Speech carries rich information such as tone, tempo, and pauses, which can be indicative of a person’s emotional state. Moreover, speech-based systems enable real-time monitoring, making them suitable for continuous assessment of mental health. Notably, Dubagunta et al. (2019) \cite{Dubagunta2019} leveraged the combination of prior knowledge-based signal processing methods and CNNs to detect depression from voice source-related information. Their study demonstrated that neurophysiological changes during depression, which affect laryngeal control, can be effectively detected using this approach, offering a promising direction for future research.

DepAudioNet represents a seminal contribution to the field of automated depression detection through audio analysis. Introduced by Ma et al. (2016) \cite{Ma2016DepAudioNet}, DepAudioNet addresses the limitations of traditional methods that predominantly rely on hand-engineered features. It automates the feature extraction and classification processes using a combination of convolutional and recurrent neural networks. The superiority of automatic feature extraction over hand-engineered features, particularly for tasks focused on paralinguistics such as speech-based depression detection, has been well documented in the literature \cite{ioannides23_interspeech}. The convolutional layers of DepAudioNet excel at capturing spectral features from audio signals \cite{Ameta2023}, while the recurrent layers, typically LSTM networks, analyze temporal relationships within the speech data \cite{Shewalkar2019}. This dual approach enables the model to effectively process and interpret the complex, non-linear relationships inherent in human speech, which are often indicative of affective states such as depression.

In recent advancements, attention mechanisms have been increasingly integrated into speech-based depression detection models to enhance performance and interpretability. Zhao et al. (2015) \cite{Zhao2015} introduced a Hierarchical Attention Transfer Network (HATN) that applies hierarchical attention autoencoders to transfer attention from a source task, such as speech recognition, to a depression detection system. This approach significantly improved the efficiency of depression severity diagnosis. However, even more sophisticated attention mechanisms have emerged.

The introduction of the Density Adaptive Attention Mechanism (DAAM) marked a pivotal development in this domain. Ioannides et al. (2024) \cite{Ioannides2024} proposed the multi-head DAAM within a Density Adaptive Transformer (DAT), presenting a probabilistic attention framework that integrates learnable mean offset and variance parameters within a multi-headed structure. This framework allows for dynamic recalibration of feature significance across diverse modalities, including speech, particularly addressing the non-stationary nature of speech data in depression detection. The DAAM framework has demonstrated substantial performance improvements, with accuracy gains of up to 20\% over existing state-of-the-art attention models. The ability of DAAM to dynamically re-calibrate features across different genders, as explored by Koudounas et al., further underlines its versatility and effectiveness in handling complex, real-world data.

The combination of DepAudioNet and DAAM, as implemented in the current work, leverages these advancements to create a more robust and explainable model for depression detection. By integrating multi-head DAAM into the existing DepAudioNet architecture, this hybrid model not only improves classification performance but also provides a more transparent decision-making process, crucial for clinical applications.

\begin{figure*}[ht!]
\centering
\includegraphics[width=\textwidth]{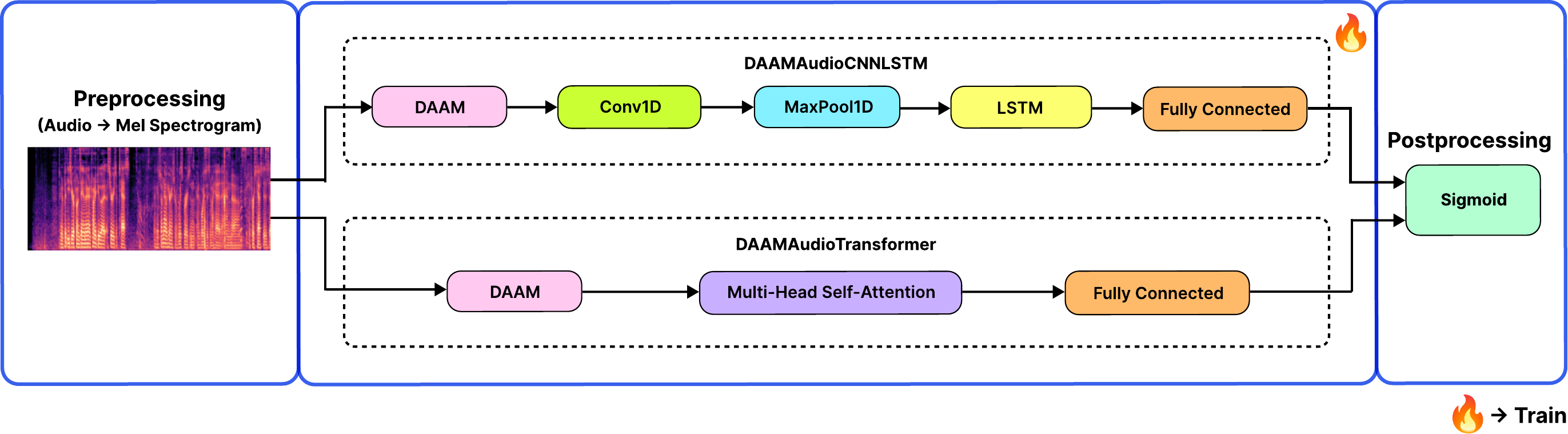}
  \caption{The two proposed network architectures featuring a DAAM module positioned after the input Mel Spectrogram and before the first layer to refine and modulate the feature extraction process.}
    \label{fig:network_architecture}
\end{figure*}

\section{Methods}

Our proposed network architectures, which are illustrated in Figure~\ref{fig:network_architecture}, are designed to process audio data for the detection of depression, utilizing a combination of attention mechanisms and deep learning models tailored for audio feature extraction.

The processing pipeline for both models begins with raw audio input, which is transformed into a Mel Spectrogram. Using a Hanning window, mel-spectrogram features are extracted with a Mel filterbank that includes 40 frequency bins, a window length of \(w = 1024\), and a hop length of \(h = 512\). These features are standardized through z-normalization, yielding adjusted features represented as \(\tilde{x}\), where \(\tilde{x} = (x - \mu) / \sigma\). Here, \(x\) is the original feature from the audio input, and \(\mu\) and \(\sigma\) denote the mean and standard deviation, respectively, calculated from the features of each distinct audio recording. This normalization aligns with the methods used by Ma et al. \cite{Ma2016DepAudioNet} and Bailey et al. \cite{Bailey2020GenderBI}, ensuring consistency and mitigating class imbalance by cropping features to align with the duration of the shortest audio track. Random subsampling is applied to non-depressed (ND) instances to balance the dataset, which is then segmented into temporal portions of length \(N_{\text{seg}} = 120\) for further processing.

\subsection{DAAMAudioCNNLSTM}
The DAAMAudioCNNLSTM model incorporates a Gaussian-based Attention Augmented Module (DAAM) before the first layer, initialized with 4 attention heads, each utilizing 16 or 24 Gaussian distributions. This mechanism focuses on enhancing the model's performance by weighting important spatio-temporal features within the Mel Frequency Cepstral Coefficients (MFCCs). The DAAM modifies the input tensor \(X\) to \(X'\), computed as:

\begin{equation}
X' = X \cdot \prod_{i=1}^{N} \frac{e^{-\frac{(X - (\mu + \bm{\delta_i}))^2}{2 \bm{\sigma_i}^2}}}{\sqrt{2 \pi \bm{\sigma_i}^2}}
\end{equation}

where \(N\) represents the number of Gaussian distributions in the mixture, \(\mu\) is the mean of \(X\), and \(\bm{\delta_i}\) and \(\bm{\sigma_i}\) are learnable parameters that adjust the mean and standard deviation for each Gaussian component.

The network architecture proceeds with a convolutional layer (kernel size: \(40 \times 3\), stride: \(1 \times 1\), padding: \(0 \times 1\)) producing a feature map of depth \(D=128\) and length \(L=120\), followed by max pooling (kernel size: 3, stride: 3, padding: 0) to reduce the length to \(L=40\). An LSTM layer with 3 layers of 128 hidden units captures temporal dependencies, leading to a fully connected layer that outputs the final prediction.

\subsection{DAAMAudioTransformer}
The DAAMAudioTransformer model enhances the processing of audio data with a CustomAttention mechanism, specifically designed for this task. Similar to DAAMAudioCNNLSTM, it starts with a Density-based attention mechanism, initialized with 4 attention heads, each using 24 Gaussian distributions. This attention mechanism selectively focuses on critical frequency bin features within the MFCCs, improving the capture of spatio-temporal characteristics.

Following the attention block, the DAAMAudioTransformer employs a series of transformer encoder layers to process the attended features. Each encoder layer comprises a multi-head self-attention mechanism and a feed-forward network (FFN), with dropout applied for regularization (dropout rate: 0.1). The transformer encoder is configured with a model dimensionality \(d_{\text{model}} = 120\), 4 attention heads, and a feed-forward dimension \(d_{\text{feedforward}} = 2048\). The output from the transformer encoder is passed through a fully connected layer with a sigmoid activation function to produce the final binary classification output, predicting the probability of depression.

\subsection{Training and Optimization}
Both models are trained using the Adam optimizer \cite{adam}, with an initial learning rate of 0.001 and a decay rate of 0.9 applied at intervals defined by \(\lambda_{\text{epoch}} = 2\). The binary cross-entropy loss (BCELoss) is employed as the training objective, quantifying the difference between the predicted probabilities and the actual binary labels. This loss function is well-suited for binary classification tasks, driving the optimization process for both models.

In summary, the DAAMAudioCNNLSTM and DAAMAudioTransformer models leverage advanced attention mechanisms and deep learning architectures to capture both local and global dependencies in audio data, making them potent tools for depression detection in speech.

\section{Results and Discussion}

\subsection{Dataset}

The Distress Analysis Interview Corpus - Wizard of Oz (DAIC-WOZ) dataset plays a central role in the development of automated methods for the detection and analysis of psychological disorders, particularly depression. Developed as part of the DARPA Detection and Computational Analysis of Psychological Signals (DCAPS) program, this dataset is designed to advance the understanding and detection of psychological stress signals with a focus on depression \cite{Gratch2014DCAPS, Ringeval2019}. We have received approval to use this dataset from the University of Southern California (USC) Institute of Creative Technologies.

The DAIC-WOZ dataset consists of 189 sessions, each corresponding to an interview between a participant and a virtual interviewer named ``Ellie." These interviews were conducted in a controlled environment, with Ellie being operated by a researcher in a separate room \cite{DeVault2014}. The interviews, varying in length from 7 to 35 minutes, provide a comprehensive multimodal dataset, including audio recordings, transcribed text, facial expressions, and physiological signals. For privacy reasons, the raw visual data has not been disclosed. Instead, the dataset provides visual features extracted using the OpenFace framework and the FACET toolbox, along with raw audio files sampled at 16 kHz \cite{Bailey2021}. For the purposes of this work, the focus is primarily on the audio component of the dataset, which offers valuable insights into the verbal and paraverbal aspects of communication that are indicative of depressive symptoms.

The training segment of the dataset comprises 107 files, detailed as follows: 27 files from females without depression (ND) and 17 from females with depression (D), contributing to a female total of 44 files (41\%); for males, 49 ND and 14 D files make up a total of 63 files (59\%), leading to an overall distribution of 76 ND (71\%) and 31 D (29\%) across genders. The validation segment includes 35 files, split into 23 ND and 12 D, to facilitate machine learning applications. The average interview time across all sessions is 956.33 seconds ($\sim$15.94 minutes), with a standard deviation of 269.96 seconds ($\sim$4.50 minutes). The distribution of interview times ranges from 414.80 seconds ($\sim$6.91 minutes) to 1966.20 seconds ($\sim$32.77 minutes), illustrating the variability in session durations.

A critical feature of the DAIC-WOZ dataset is its imbalance in terms of participants' level of depression. The dataset is heavily biased towards participants with PHQ-8 scores below 10, indicating no to mild depression. For instance, the total time allocated to interviews with participants scoring below 10 is 2,121.14 minutes ($\sim$35.35 hours), while participants scoring 10 or higher account for only 891.30 minutes ($\sim$14.86 hours). This imbalance underscores the necessity of careful handling during model development and evaluation to prevent bias towards the overrepresented class.

The dataset is divided into three subsets: Training, Testing, and Development, essential for the development and evaluation of predictive models for depression detection. The training split consists of 107 participants, categorized into depressed (D) and non-depressed (ND), with a total interview time of 454.00 minutes ($\sim$7.57 hours) for D participants and 1,160.22 minutes ($\sim$19.34 hours) for ND participants. The test split comprises 47 participants, with a total interview time of 198.92 minutes ($\sim$3.32 hours) for D participants and 598.28 minutes ($\sim$9.97 hours) for ND participants. The development split consists of 35 participants, with a total interview time of 238.38 minutes ($\sim$3.97 hours) for D participants and 362.64 minutes ($\sim$6.04 hours) for ND participants. In this study, we assess the performance of our models on the validation portion of the dataset, aligning our methodology with established practices in the literature.

\begin{table}[h]
\centering
\begin{tabular}{lccc}
\toprule
\textbf{Split}       & \textbf{D Time (min)} & \textbf{ND Time (min)} & \textbf{Total Time (min)} \\
\toprule
Train       & 454.00       & 1,160.22      & 1,614.22         \\
Test        & 198.92       & 598.28        & 797.20           \\
Development & 238.38       & 362.64        & 601.02           \\
\midrule
\textbf{Total} & 891.30     & 2,121.14      & 3,012.44         \\
\bottomrule

\end{tabular}
\caption{Dataset splits in terms of time.}
\label{tab:dataset_splits_time}
\end{table}

\begin{table}[h]
\centering
\begin{tabular}{lccc}
\toprule
\textbf{Split}       & \textbf{D Participants} & \textbf{ND Participants} & \textbf{Total Participants} \\
\toprule
Train       & 30             & 77              & 107               \\
Test        & 12             & 35              & 47                \\
Development & 12             & 23              & 35                \\
\midrule
\textbf{Total} & 54          & 135             & 189               \\
\bottomrule
\end{tabular}
\caption{Dataset participant distribution.}
\label{tab:dataset_participant_distribution}
\end{table}

It is important to note that the dataset contains certain known errors and special cases that need to be considered during analysis. Every file contains pre-interview interactions that must be excised to isolate the interview content. Certain sessions (e.g., 373 and 444) include lengthy interruptions, while others (e.g., 451, 458, and 480) lack the virtual agent's transcriptions. Transcription and audio synchronization issues are present in sessions 318, 321, 341, and 362. Additionally, a labeling error was identified in the PHQ-8 scores, where interviews with scores of 10 or above were mistakenly labeled with a binary value of 0 instead of 1. The affected interviews are listed in Table~\ref{tab:wrong_phq8_scores}.

\begin{table}[h]
\centering
\begin{tabular}{ccc}
\toprule
\textbf{Participant ID} & \textbf{PHQ-8 Score} & \textbf{Incorrect Label} \\
\toprule
320            & 11          & 0              \\
325            & 10          & 0              \\
335            & 12          & 0              \\
344            & 11          & 0              \\
352            & 10          & 0              \\
356            & 10          & 0              \\
380            & 10          & 0              \\
386            & 11          & 0              \\
409            & 10          & 0              \\
413            & 10          & 0              \\
418            & 10          & 0              \\
422            & 12          & 0              \\
433            & 10          & 0              \\
459            & 16          & 0              \\
\bottomrule
\end{tabular}
\caption{Wrong PHQ-8 scores in the dataset.}
\label{tab:wrong_phq8_scores}
\end{table}

Another concern is the gender distribution within the training split of the dataset, which is slightly male-biased with 59\% (63) being men. Among these, only 13\% (14) of the total male participants are categorized as ``D", while 16\% (17) of the female participants are categorized as ``D".

\subsection{Evaluation Metrics}
In our study, we employ the Macro F1 Score as the primary evaluation metric, which offers a more robust performance measure in the context of imbalanced datasets, such as the DAIC-WOZ dataset. The Macro F1 Score is defined as the average of the F1 Scores of each class, calculated by Equation \ref{rrr}.

\begin{equation}
    \text{Macro F1 Score} = \frac{2 \cdot \text{Precision} \cdot \text{Recall}}{\text{Precision} + \text{Recall}}
    \label{rrr}
\end{equation}

where Precision is the ratio of true positives to the sum of true positives and false positives, and Recall is the ratio of true positives to the sum of true positives and false negatives, averaged over all classes. Furthermore, we utilize a metric termed the Importance Factor (IF), as introduced by Ioannides et al. \cite{Ioannides2024}, which is derived from the Density Attention (DA) weights in the attention module. The IF for each feature is calculated using Equation \ref{lll}.

\begin{equation}
    \text{IF}_{ij} = \frac{\text{GA}_{ij} - \min(\text{GA})}{\max(\text{GA}) - \min(\text{GA})}
    \label{lll}
\end{equation}

The IF values range from 0 to 1 (inclusively), with higher values denoting features of increased significance in the model's decision-making process as shown by Ioannides et al. \cite{Ioannides2024}. To visually depict the importance of features, we construct heatmaps based on the IF. These heatmaps are created by averaging the Density attention maps obtained during the validation phase and subsequently applying the IF metric as shown in Equation \ref{avg}.

\begin{equation}
    \text{Heatmap} = \frac{1}{N} \sum_{i=1}^{N} \text{GA}_i
    \label{avg}
\end{equation}

The resulting graphical representations serve to highlight the feature importance as determined by the model.

\begin{figure*}[t]
    \centering
    \captionsetup[subfigure]{labelfont=normalfont,textfont=normalfont}
    
    \begin{subfigure}[t]{0.45\textwidth}
        \centering
        \includegraphics[width=\linewidth, trim=0 0 0 0, clip]{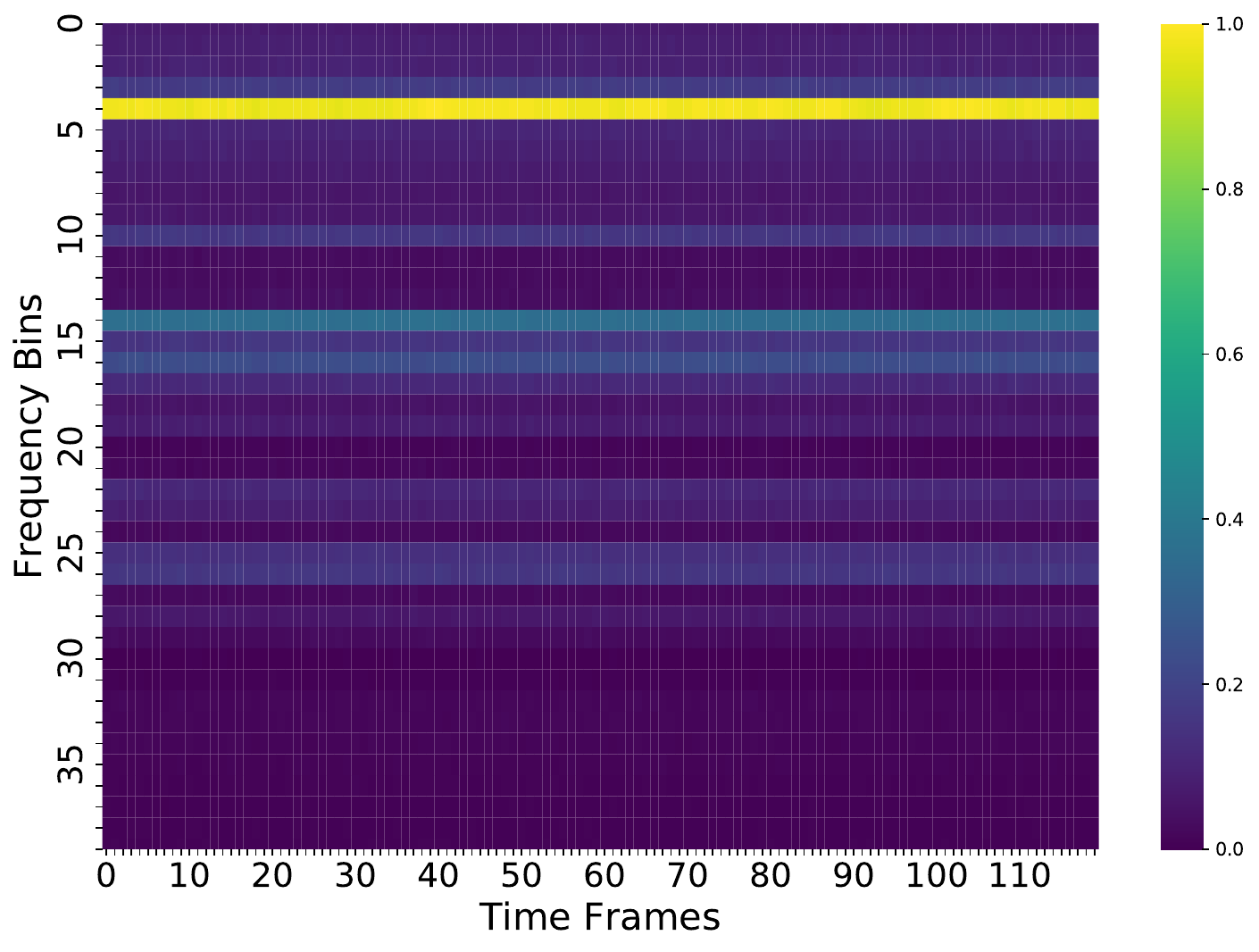}
        \caption{Mel spectrogram Density Attention Importance Factor values for $g=16$ in DAAMAudioCNNLSTM.}
        \label{fig:mel16}
    \end{subfigure}
    \hfill
    \begin{subfigure}[t]{0.45\textwidth}
        \centering
        \includegraphics[width=\linewidth, trim=0 0 0 0, clip]{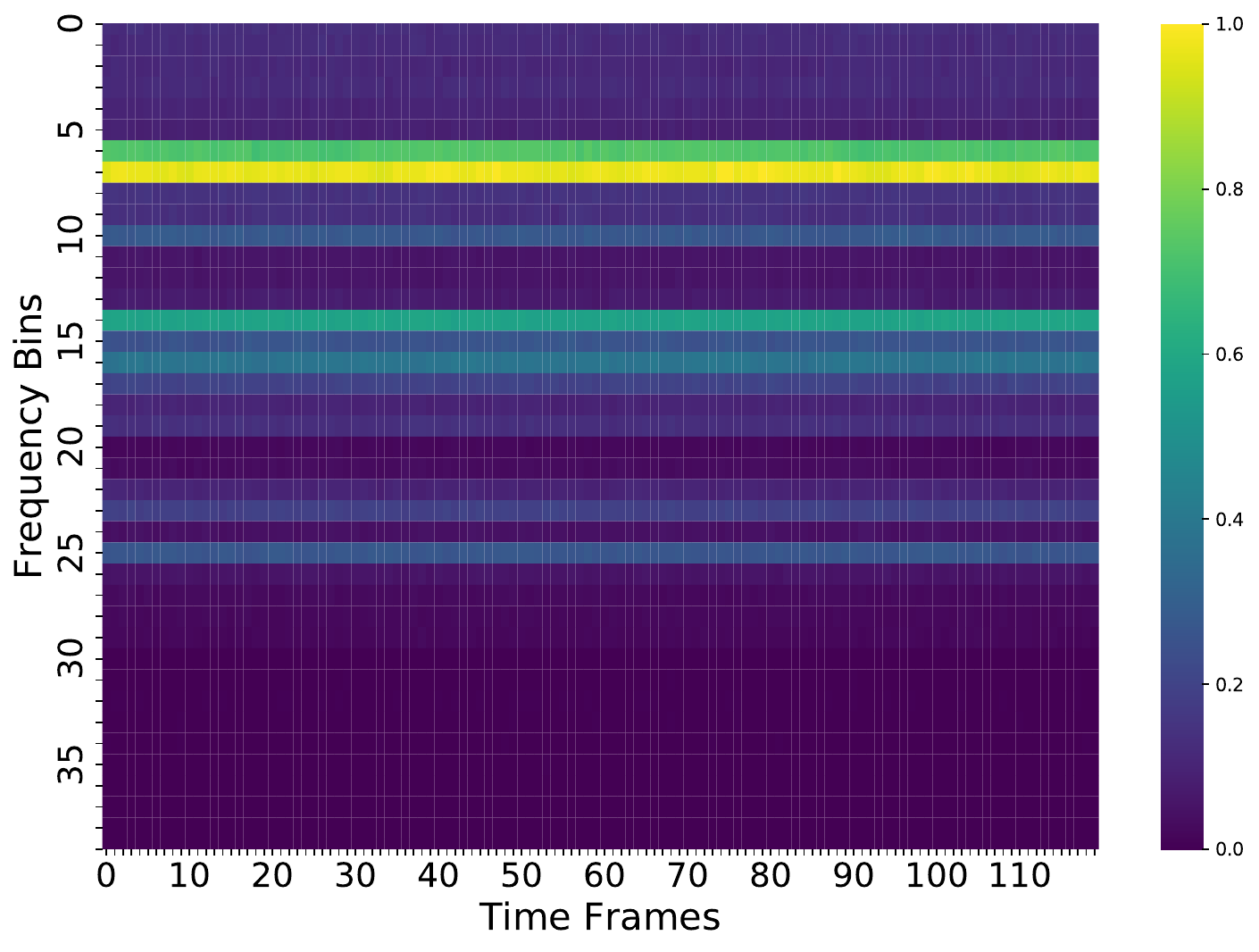}
        \caption{Mel spectrogram Density Attention Importance Factor values for $g=24$ in DAAMAudioCNNLSTM.}
        \label{fig:mel24}
    \end{subfigure}

    \vspace{0.2cm}

    \begin{subfigure}[t]{0.45\textwidth}
        \centering
        \includegraphics[width=\linewidth, trim=0 0 0 0, clip]{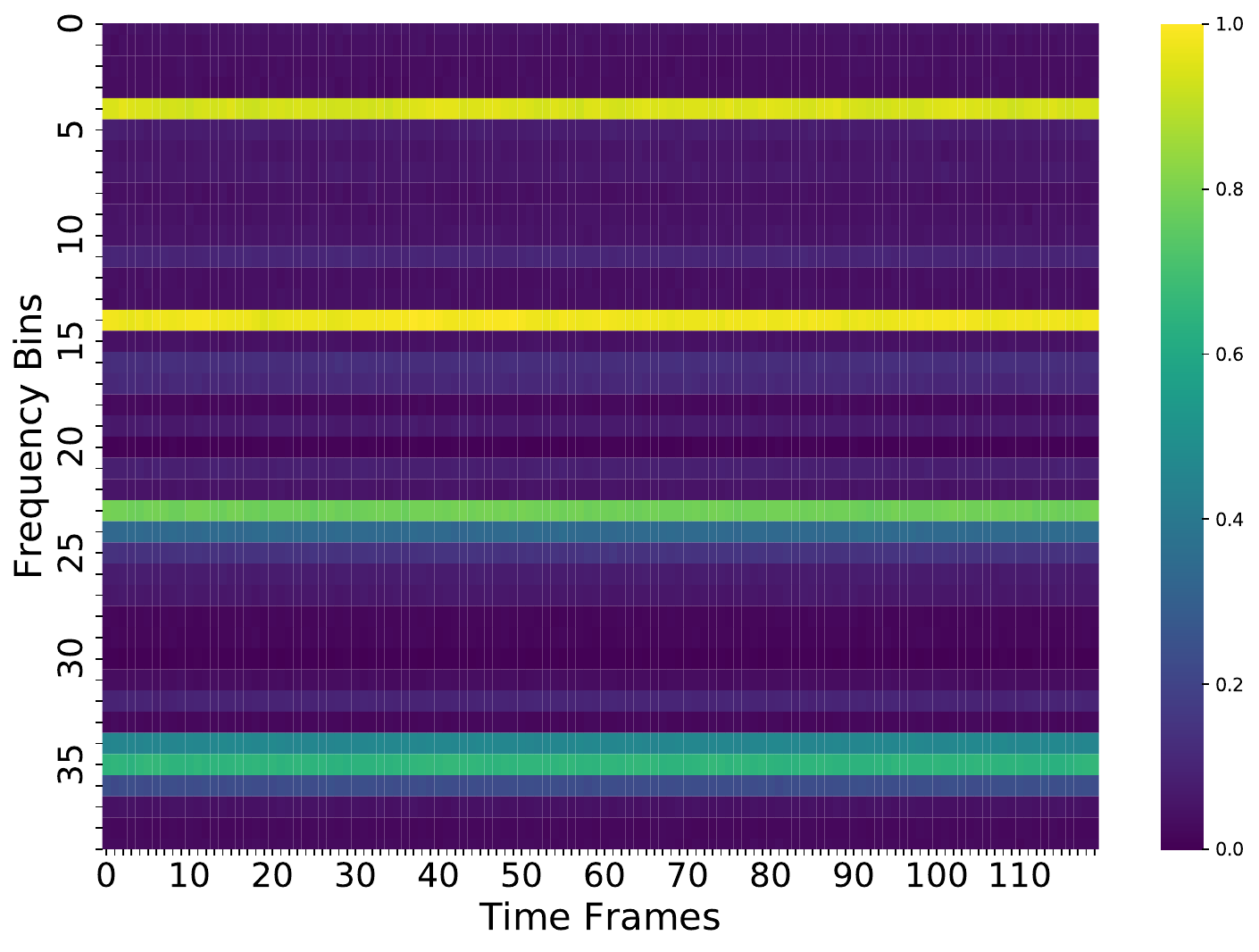}
        \caption{Mel spectrogram Density Attention Importance Factor values for $g=16$ in DAAMAudioTransformer.}
        \label{fig:customattentiontransformer_g16}
    \end{subfigure}
    \hfill
    \begin{subfigure}[t]{0.45\textwidth}
        \centering
        \includegraphics[width=\linewidth, trim=0 0 0 0, clip]{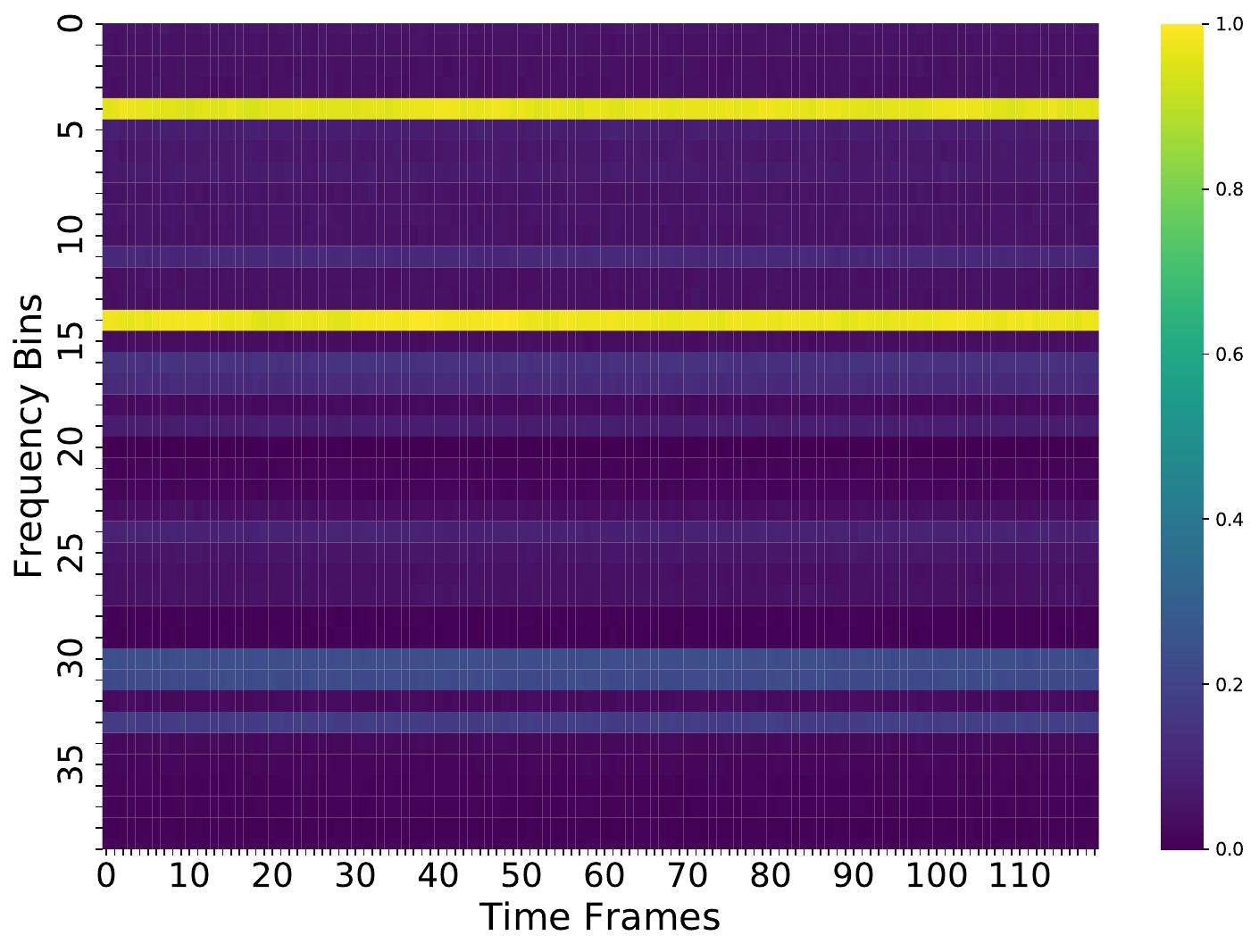}
        \caption{Mel spectrogram Density Attention Importance Factor values for $g=24$ in DAAMAudioTransformer.}
        \label{fig:customattentiontransformer_g24}
    \end{subfigure}
    
    \caption{Comparison of Mel spectrogram Density Attention Importance Factor values for different models and configurations.}
    \label{fig:comparison}
\end{figure*}

\begin{table*}[htbp]
    \centering
    \caption{Performance of Models on Speech Depression Detection Using Only Depression Labels.}
    \scriptsize 
    \resizebox{0.75\textwidth}{!}{      
        \begin{tabular}{lcccc}
        \toprule
        \textbf{Model} & \textbf{F1 (ND)} & \textbf{F1 (D)} & \textbf{F1 (Avg.)} & \textbf{Number of Params} \\
        \midrule
        DepAudioNet \cite{Ma2016DepAudioNet} & 0.700 & 0.520 & 0.610 & 280K \\
        DepAudioNet$^*$(Mel Spectrogram) \cite{Bailey2020GenderBI} & 0.740 & 0.539 & 0.634 & 280K \\
        FVTC-CNN \cite{Huang2020VocalTractCoordination} & 0.460 & 0.820 & 0.640 & $\sim$19.1K \\
        DepAudioNet$^*$(Raw Audio) \cite{Bailey2020GenderBI} & 0.796 & 0.520 & 0.658 & 280K \\
        SpeechFormer \cite{chen2022speechformerhierarchicalefficientframework} & - & - & 0.694 & 33.21M \\
        \midrule
        DAAMAudioCNNLSTM ($h=4, g=16$) & 0.792 & 0.615 & 0.694 & 280K \\
        \textbf{DAAMAudioCNNLSTM} ($h=4, g=24$) & 0.815 & 0.643 & \textbf{0.702} & 280K \\
        \textbf{DAAMAudioTransformer} ($h=4, g=24$) & 0.84 & 0.6 & \textbf{0.72} & $\sim$1.1M \\
        \bottomrule
        \end{tabular}
    }    
    \label{table:results}
\end{table*}

\subsection{Current Benchmarks}
In Table \ref{table:results}, we present a benchmark comparison of various models for Speech Depression Detection, utilizing solely depression labels for training. Notably, both DAAMAudioCNNLSTM and DAAMAudioTransformer distinguish themselves through their performance metrics and architectural efficiencies.

DAAMAudioCNNLSTM, with 280K parameters, is particularly noteworthy for its lightweight architecture, which contrasts favorably with other models in the literature, such as SpeechFormer \cite{chen2022speechformerhierarchicalefficientframework}, which utilizes 33M parameters. The low parameter count of DAAMAudioCNNLSTM results in a model that is more computationally efficient, facilitating faster training and inference times. This efficiency is advantageous in scenarios with limited computational resources, translating into reduced energy consumption and lower operational costs, making it a more sustainable choice for large-scale deployment, especially in mobile and edge computing environments where power and computational capabilities are constrained.

The DAAMAudioTransformer, on the other hand, while more parameter-heavy with 1.1M parameters, offers a more sophisticated approach to feature extraction and data processing. Its architecture integrates custom attention mechanisms with transformer encoder layers, enabling it to capture both local and global dependencies in audio data effectively. Despite its higher parameter count, the DAAMAudioTransformer remains significantly smaller than other state-of-the-art models like WavLM, which contains 316M parameters \cite{chensWavlm}.

When comparing performance metrics, DAAMAudioCNNLSTM achieved an F1 score of 0.815 for ND classifications and 0.643 for D classifications, leading to a macro F1 score of 0.702. This score underscores its strong ability to balance precision and recall across different categories. The DAAMAudioTransformer, however, slightly outperforms DAAMAudioCNNLSTM in the ND category, with an F1 score of 0.84 for ND classifications and 0.6 for D classifications, resulting in a higher macro F1 score of 0.72. This makes DAAMAudioTransformer particularly effective in detecting non-depressive states, which is a significant achievement considering its more complex architecture.

The lower parameter count in DAAMAudioCNNLSTM implies a reduced propensity for overfitting, adhering to the principle of Occam's razor \cite{BargagliStoffi2022}. This enhances the model's generalizability to unseen data, a critical factor for real-world applications where data diversity can be substantial. In contrast, while the DAAMAudioTransformer's larger size does introduce more computational demands, it also allows for a more nuanced understanding of the input data, particularly in capturing a broader spectrum of frequency bins, which contributes to its superior performance in ND detection.

{
\fontsize{8pt}{10pt}\selectfont
\begin{table}[htbp]
    \centering
    \caption{Summary of Mean Offsets and Variances Across Heads for $h=4$ and $g=16$ in DAAMAudioCNNLSTM.}
\resizebox{0.85\columnwidth}{!}{    
    \begin{tabular}{ccc}
    \toprule
    \textbf{Head} & \textbf{Mean Offsets} $\boldsymbol{\delta}$ \textbf{(Min, Max)} & $\boldsymbol{\sigma^{2}}$ \textbf{(Min, Max)} \\
    \midrule
    0 & (-0.04, -0.04) & ($4 \times 10^{-8}$, 1.59) \\
    1 & (-0.04, -0.04) & ($10^{-4}$, 5.43) \\
    2 & (0.02, 0.02) & ($10^{-8}$, 99.60) \\
    3 & (-0.03, -0.01) & ($10^{-4}$, 33.64) \\
    \bottomrule
    \end{tabular}
}    
    \label{g16}
\end{table}
}

{
\fontsize{8pt}{10pt}\selectfont
\begin{table}[htbp]
    \centering
    \caption{Summary of Mean Offsets and Variances Across Heads for $h=4$ and $g=24$ in DAAMAudioCNNLSTM.}
\resizebox{0.85\columnwidth}{!}{      
    \begin{tabular}{ccc}
    \toprule
    \textbf{Head} & \textbf{Mean Offsets} $\boldsymbol{\delta}$ \textbf{(Min, Max)} & $\boldsymbol{\sigma^{2}}$ \textbf{(Min, Max)} \\
    \midrule
    0 & (0.02, 0.02) & ($10^{-4}$, 5.29) \\
    1 & (-0.04, -0.04) & ($10^{-4}$, 12.18) \\
    2 & (0.01, 0.03) & ($10^{-8}$, 14.14) \\
    3 & (-0.01, 0.02) & ($10^{-8}$, 19.00) \\
    \bottomrule
    \end{tabular}
}    
    \label{g24}
\end{table}
}
{
\fontsize{8pt}{10pt}\selectfont
\begin{table}[htbp]
    \centering
    \caption{Summary of Mean Offsets and Variances Across Heads for $h=4$ and $g=24$ in DAAMAudioTransformer.}
\resizebox{0.85\columnwidth}{!}{      
    \begin{tabular}{ccc}
    \toprule
    \textbf{Head} & \textbf{Mean Offsets} $\boldsymbol{\delta}$ \textbf{(Min, Max)} & $\boldsymbol{\sigma^{2}}$ \textbf{(Min, Max)} \\
    \midrule
    0 & (0.0, 0.0) & (0.001346, 4.067) \\
    1 & (0.0, 0.0) & (0.003184, 3.139) \\
    2 & (0.0, 0.0) & (0.000101, 11.445) \\
    3 & (-0.01096, -0.01070) & (0.002957, 7.206) \\
    \bottomrule
    \end{tabular}
}    
    \label{g24_C}
\end{table}
}

Both models employ attention mechanisms that are crucial for their performance in speech depression detection. DAAMAudioCNNLSTM utilizes a DAAM module that distributes focus across four attention heads, each operating on a subset of the 40 Mel frequency bins derived from the speech signal. The attention distribution is quantified by two parameters per head: the mean offset (\(\delta\)) and the standard deviation (\(\sigma\)). For DAAMAudioCNNLSTM, the mean offset values are confined within a narrow range (\(\delta \in [-0.04, 0.02]\)), indicating consistent adjustments to the attention weights across all heads.

DAAMAudioTransformer also displays a detailed attention mechanism. The analysis of its attention heads reveals that while the mean offsets are consistently centered at 0.0 for some heads, there is a broader variance across others, particularly in Heads 2 and 3. This broader variance allows for a wider spread of focus, making the model highly adaptable to changes in the audio spectrum. The variance ($\sigma^{2}$) values range from $[0.001346, 4.067]$ for Head 0 to $[0.000101, 11.445]$ for Head 2, suggesting an adaptable focus that can adjust to varying audio features.

In terms of classification performance, the DAAMAudioTransformer model shows a balanced accuracy of 0.7065, with an accuracy of 0.913 for the ND category and 0.5 for the D category. The confusion matrix for DAAMAudioTransformer indicates that the model made fewer errors in predicting ND states compared to D states, highlighting its strength in recognizing non-depressive features. The model's ability to handle both high and low frequencies effectively is further demonstrated by the IF analysis, where Bin 14 (141.74 Hz - 251.84 Hz), Bin 4 (856.36 Hz - 1059.93 Hz), and Bin 30 (2006.13 Hz - 2360.09 Hz) were assigned the highest importance, reflecting a comprehensive focus across a wide range of frequencies.

Both models were trained with 4 attention heads and 24 Gaussian distributions per head. The DAAMAudioCNNLSTM model primarily focuses on higher frequencies, particularly those associated with vocal energy and the naturalness of speech, such as the 2006.13 Hz - 2360.09 Hz range. This focus suggests its effectiveness in detecting subtle features in speech that contribute to a more natural and pleasant listening experience.

Conversely, the DAAMAudioTransformer targets a broader spectrum that includes both lower and mid-range frequencies, specifically focusing on the 141.74 Hz - 251.84 Hz range and the 856.36 Hz - 1059.93 Hz range. These frequencies are more central to the average male and female voice ranges, capturing essential prosodic elements and vocal characteristics that are critical for detecting emotional states like depression.

The emphasis on these specific frequency ranges is particularly significant in the context of Explainable AI (XAI). By identifying and analyzing which frequency bands are given priority by the model's attention mechanism, researchers and clinicians can gain insights into the acoustic features that are most indicative of depressive symptoms. This level of interpretability is crucial for validating the model's decisions and ensuring that it aligns with human understanding of speech patterns associated with depression.

For example, the DAAMAudioTransformer's attention to the 141.74 Hz - 251.84 Hz range corresponds to fundamental frequencies often linked to the human voice's core pitch, which can be altered in cases of depression. The attention to the 856.36 Hz - 1059.93 Hz range captures harmonics and other speech characteristics that contribute to the overall tonal quality of the voice, which may also be affected in depressive states. By focusing on these ranges, the DAAMAudioTransformer not only achieves high performance metrics but also provides a pathway for explainability, helping experts understand why certain predictions are made, thus making the model more trustworthy in clinical applications.

This broader focus likely contributes to its superior performance in detecting non-depressive states, as it allows the model to capture a wider array of vocal features that are indicative of normal emotional states. The ability to discern between these frequency bands and correlate them with depression-related features enhances the model's robustness and its utility in real-world applications where interpretability and accuracy are paramount.

\section{Conclusion}
This work introduces two novel networks --DAAMAudioCNNLSTM and DAAMAudioTransformer—designed for the detection of depression directly from speech signals. DAAMAudioCNNLSTM combines a CNN-LSTM architecture with a lightweight Density Adaptive Attention Mechanism, achieving a state-of-the-art F1 macro score of 0.702 on the DAIC-WOZ dataset. Similarly, DAAMAudioTransformer leverages a transformer-based architecture with custom Density attention mechanisms, capturing both local and global dependencies in the audio data, resulting in a outperforming F1 macro score of 0.72 on the same Dataset. Importantly, both models achieve these results without relying on supplementary data labels beyond depression states.

Both DAAMAudioCNNLSTM and DAAMAudioTransformer excel in explainability, efficiency in processing speech signals, and transparency in their decision-making processes, making them strong candidates for clinical adoption. These models not only set new benchmarks for speech-based mental health assessment but also pave the way for future research into acoustic biomarkers of depression through explainable artificial intelligence. Their potential for clinical application highlights significant advancements in automated depression diagnostics and in the broader domain of speech-related health.
\bibliographystyle{IEEEtran}
\bibliography{mybib}

\end{document}